\documentclass[twocolumn,english,showkeys,showpacs,superscriptaddress]{revtex4-1}
\usepackage[LGR,T1]{fontenc}
\usepackage[latin9]{inputenc}
\usepackage{amsmath}
\usepackage{amssymb}
\usepackage{graphicx}

\makeatletter

\ProvideTextCommand{\~}{LGR}[1]{\char126#1}

\usepackage[normalem]{ulem}

\usepackage{color}

\usepackage{ulem}

\newcommand{\stkout}[1]{\ifmmode\text{\sout{\ensuremath{#1}}}\else\sout{#1}\fi}

\makeatother

\usepackage{babel}
\begin{document}
\title{Theoretical Study on Superradiant Raman Scattering with Rubidium Atoms in An Optical Cavity }

\author{Huihui Yu}
\address{Henan Key Laboratory of Diamond Optoelectronic Materials and Devices, Key Laboratory of Material Physics Ministry of Education, School of Physics and Microelectronics, Zhengzhou University, Daxue Road 75, Zhengzhou 450052, China}


\author{Yuan Zhang}
\email{yzhuaudipc@zzu.edu.cn}
\address{Henan Key Laboratory of Diamond Optoelectronic Materials and Devices, Key Laboratory of Material Physics Ministry of Education, School of Physics and Microelectronics, Zhengzhou University, Daxue Road 75, Zhengzhou 450052, China}
\address{Institute of Quantum Materials and Physics, Henan Academy of Sciences, Zhengzhou 450046, China}

\author{Gang Chen}
\address{Henan Key Laboratory of Diamond Optoelectronic Materials and Devices, Key Laboratory of Material Physics Ministry of Education, School of Physics and Microelectronics, Zhengzhou University, Daxue Road 75, Zhengzhou 450052, China}
\address{Institute of Quantum Materials and Physics, Henan Academy of Sciences, Zhengzhou 450046, China}


\author{Chongxin Shan}
\email{cxshan@zzu.edu.cn}
\address{Henan Key Laboratory of Diamond Optoelectronic Materials and Devices, Key Laboratory of Material Physics Ministry of Education, School of Physics and Microelectronics, Zhengzhou University, Daxue Road 75, Zhengzhou 450052, China}
\address{Institute of Quantum Materials and Physics, Henan Academy of Sciences, Zhengzhou 450046, China}


\begin{abstract}
Superradiant Raman scattering of Rubidium atoms has been explored in the experiment [Nature 484, 78 (2012)] to prove the concept of the superradiant laser,  which attracts significant attentions in quantum metrology due to the expected ultra-narrow linewidth down to millihertz. To better understand  the physics involved in this experiment, we have developed a quantum master equation theory by treating the Rubidium atoms as three-level systems, and coupling them with a dressed laser and an optical cavity. Our simulations show  different superradiant Raman scattering pulses for the systems within the crossover and strong coupling regime, and the shifted and broader spectrum of the steady-state Raman scattering. Thus, our studies provide a unified view on the superradiant Raman scattering pulses, and an alternative explanation to the broad spectrum of the steady-state Raman scattering, as observed in the experiment. In future,  our theory can be readily applied to study other interesting phenomena relying on the superradiant Raman scattering, such as  magnetic field sensing,  real-time tracking of quantum phase,  Dicke phase transition of non-equilibrium dynamics and so on. 
\end{abstract}

\maketitle

\section{Introduction} 

Superradiance, i.e. collective spontaneous emission of quantum emitters,  was firstly proposed by R. H. Dicke in 1954~\citep{RHDicke1954}, and was then studied extensively in theories and experiments  in 1980s~\citep{AVAndreev1980}. Because this effect was often studied with the quantum emitters under the pulsed light illumination, it was  considered as a transient phenomenon following the collective decay of the emitters.  However, in 2009, D. Meiser et al. suggested that such a collective decay can be compensated by an incoherent atomic pumping, and predicted that the resulted superradiant laser can  have a ultra-narrow linewidth down to millihertz for optical lattice clock  systems~\citep{DMeiser2009}. Due to the application potential in quantum metrology, the superradiant laser has been studied extensively in both theories~\citep{DATieri2017,KDebnath2018,YZhang2021} and experiments~\citep{JDBohnet2012,MANorcia2016,MANorcia2016-1} thereafter.  

\begin{figure}[!htp]
\begin{centering}
\includegraphics[scale=0.45]{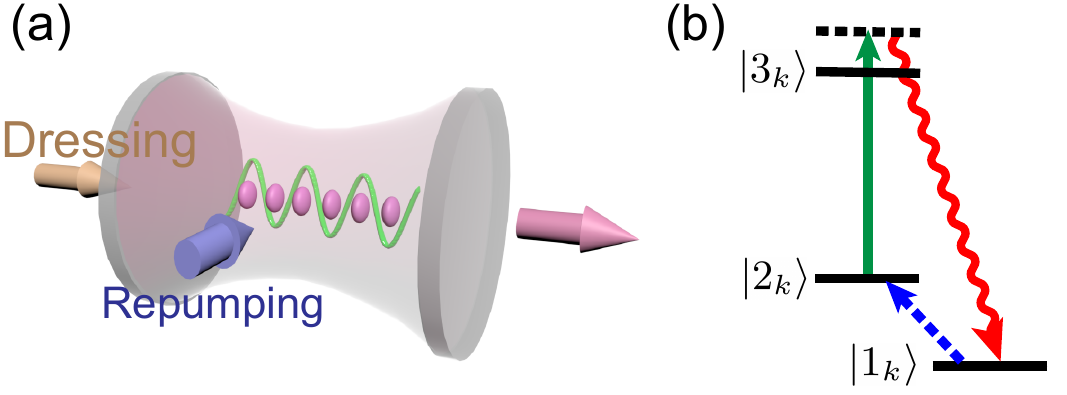}
\par\end{centering}
\caption{ \label{fig:sys} System and energy diagram.  Panel (a) shows tens of thousands of Rubidium-87 atoms trapped in an optical lattice inside an optical cavity,  a dressed (re-pumping) laser transmitting (transecting) the optical cavity, and the superradiant Raman signal out of the cavity.  Panel (b) displays a simplified energy diagram, where the dressed laser excites the atoms from the hyper-fine ground level $\left|2_k\right\rangle=\left|5^2 S_{1/2}, F=2, m_f =-2 \right\rangle$ to the virtual level (dashed line) slightly above the excited level  $\left|3_k\right\rangle=\left|5^2 P_{1/2}, F=2, m_f =-2 \right\rangle$ (green arrow), and the coupling with the optical cavity de-excites the atoms to the hyper-fine level $\left|1_k\right\rangle=\left|5^2 S_{1/2}, F=1, m_f =-1 \right\rangle$(red wavy arrow), producing superradiant Raman scattering pulses.  If the atoms are re-pumped to the higher hyper-fine ground level (blue dashed arrow), the superradiant Raman scattering can be maintained. }
\end{figure}

Among the experiments on the superradiant laser,  the one by J. G. Bohnet  et al, was often viewed as a proof-of-concept of such an effect~\citep{JDBohnet2012}, although it relied on  Raman transition of Rubidium atoms (Fig.~\ref{fig:sys})  rather than  real optical clock transitions.  To better understand the physics involved,  in this article, we develop a quantum master equation theory by treating the atoms as three-level systems and coupling them with a dressed laser and an optical cavity.  By solving the master equation with cumulant mean-field approach, we simulate the systems with tens of thousands of Rubidium atoms. 

Our calculations show the different superradiant Raman scattering pulses for the system in the crossover and strong coupling regime, and the shifted and broader Raman scattering spectrum for the system at steady-state. Thus, our results provide a unified view on the superradiant Raman pulses, and an alternative explanation on the broad spectrum, as observed in the experiments. In future, our theory can be readily applied to study other interesting phenomena relying on the superradiant Raman scattering, such as magnetic field sensing~\citep{JMWeiner2012}, real-time track of quantum phase~\citep{AShankar2019}, Dicke phase transition of non-equilibrium dynamics ~\citep{MPBaden2019}, and so on. 

The current article is organized as follows.  In the following section, we present the corresponding quantum master equation, the effective master equation after eliminating the optically excited level, and their solution with the cumulant mean-field approach. In Sec.~\ref{sec:pulses} and ~\ref{sec:continuous}, we present our numerical results on the transient and steady-state superradiant Raman scattering, respectively. In the end, we summarize our work and comment on the possible studies in the future.

\section{Quantum Master Equation} 

The system under consideration consists of tens of thousands of Rubidium-87 atoms trapped in an optical lattice inside an optical cavity (Fig.~\ref{fig:sys}a). A dressed laser transmits through the cavity, and excites the atoms from the hyper-fine ground level  to the virtual level slightly above the excited level, and the coupling with the cavity de-excites the atoms to the hyper-fine level, leading to the superradiant Raman scattering pulses (Fig.~\ref{fig:sys}b). If the atoms are also re-pumped to the higher hyper-fine ground level, the superradiant Raman scattering can be maintained. Note that the current diagram is only valid for the system within the weak or crossover coupling regime, and the excited level of the bare atoms should be replaced by the atoms-photon dressed states for the system within the strong coupling regime. 

To describe the aforementioned dynamics, we establish the following quantum master equation for the reduced density operator $\hat{\rho}$ in the interaction picture: 
\begin{align}
& \frac{\partial}{\partial t} \hat{\rho} =-\frac{i}{\hbar}[\hat{H}_{a-c}+\hat{H}_{d},\hat{\rho}] -\kappa\mathcal{D}\left[\hat{a}\right]\hat{\rho}\nonumber\\
&-\sum_{k=1}^{N}(\gamma_{31}\mathcal{D}\left[\hat{\sigma}_{k}^{13}\right]\hat{\rho} +\gamma_{12}\mathcal{D}\left[\hat{\sigma}_{k}^{21}\right]\hat{\rho}). \label{eq:qme}
\end{align}
In this equation, the Hamiltonian 
$\hat{H}_{a-c}=\hbar g_{31} [e^{i\left(\omega_{c}-\omega_{31}\right)t}\hat{a}^{\dagger} \sum_{k=1}^{N}\hat{\sigma}_{k}^{13} +e^{-i\left(\omega_{c}-\omega_{31}\right)t}\hat{a} \sum_{k=1}^{N}\hat{\sigma}_{k}^{31}]$ describes the coherent energy exchange between the atoms and the optical cavity with strength $g_{31}$. Here, the optical cavity is modeled as a quantum harmonic oscillator with a frequency $\omega_{c}$, and the photon creation operator $\hat{a}^{\dagger}$ and annihilation operator $\hat{a}$. $\hat{\sigma}_{k}^{13}$, $\hat{\sigma}_{k}^{31}$ are the lowering and raising operator associated with the  $1_k\to 3_k$ transition of frequency $\omega_{31}$. The index $k$ indicates the individual of the total $N$ atoms. In Eq.~\eqref{eq:qme}, the Hamiltonian $\hat{H}_{d}=\hbar \Omega [e^{i\left(\omega_{d}-\omega_{32}\right)t} \sum_{k=1}^{N}\sigma_{k}^{23} +e^{-i\left(\omega_{d}-\omega_{32}\right)t} \sum_{k=1}^{N}\sigma_{k}^{32} ]$ describes the driving of the atoms by a dressing laser of  frequency $\omega_{d}$ with a strength $\Omega$, where $\sigma_{k}^{23},\sigma_{k}^{32}$ are the lowering and raising operators related to the $2_k\to 3_k$ transition of frequency $\omega_{32}$. The remaining terms of Eq.~\eqref{eq:qme} describe the system dissipation with the Lindblad superoperator $\mathcal{D}\left[\hat{o}\right]\hat{\rho}=\frac{1}{2}\left(\hat{o}^{\dagger}\hat{o}\hat{\rho}+\hat{\rho}\hat{o}^{\dagger}\hat{o}\right)-\hat{o}\hat{\rho}\hat{o}^{\dagger}$
(for any operator $\hat{o}$), which includes the  photon loss of the cavity with a rate $\kappa$, the spontaneous emission and the incoherent pumping of atoms with the rates $\gamma_{31},\gamma_{12}$. The incoherent pumping can be implemented by coupling the multiple-level atoms with a sequence of laser pulses, as implemented in the experiment~\citep{JDBohnet2012}. Here, we consider the rather complex process phenomenologically with a dissipative process, which is the inverse of the spontaneous emission and transfers incoherently the population from the lower to higher hyper-fine ground level. For the sake of simplicity, we have also ignored other possible decay and dephasing processes, but emphasize that they can be easily incorporated into the master equation. The Raman scattering spectrum can be calculated from the expression ${\rm S}(\omega)=2\kappa{\rm Re}[\int d\tau e^{-i\omega \tau} \left \langle \hat{a}^{\dagger}(\tau) \hat{a}(0)\right \rangle]$, where the equation for the two-time correlation function (the integral kernel) can be obtained by applying quantum regression theorem~\citep{DMeiser2009}.

The experiment of J. G. Bohnet  et al. is motivated by that the  dynamics for the current system in weak coupling regime   is similar to the superradiant laser system. To see this point, we eliminate adiabatically the excited level by defining the slowly varying term $\hat{\bar{\sigma}}_{k}^{13} \equiv e^{i(\omega_l - \omega_{32})t} \hat{\sigma}_{k}^{13}$, and deriving the corresponding Heisenberg equation   $\partial_t \hat{\bar{\sigma}}_{k}^{13} \approx i(\omega_d - \omega_{32} +i \gamma_{31}/2) \hat{\bar{\sigma}}_k^{13} - i\Omega \hat{\sigma}_k^{12}$. Then, we solve  this equation at steady-state to obtain  $\hat{\bar{\sigma}}_k^{13} \approx [\Omega/(\omega_d - \omega_{32} +i \gamma_{31}/2)] \hat{\sigma}_k^{12}$, and insert this solution into Eq.~\eqref{eq:qme} to obtain an effective quantum master equation:
\begin{align}
& \partial_{t} \hat{\rho} = -\frac{i}{\hbar} [\hat{H}'_{a-c}(t),\hat{\rho}] - \kappa \mathcal{D}[\hat{a}] \hat{\rho} \nonumber\\
&- \sum_{k=1}^N (\gamma_{21} \mathcal{D}[\hat{\sigma}_k^{12}] \hat{\rho}+\gamma_{12} \mathcal{D}\left[\hat{\sigma}_{k}^{21}\right]\hat{\rho}). \label{eq:eff-qme}
\end{align}
In this equation, the effective Hamiltonian 
$\hat{H}'_{a-c}(t) = \hbar g_{21} \left[e^{i(\omega_c - \omega_{31} -\omega_d +\omega_{32}) t} \hat{a}^\dagger  \sum_{k=1}^N \hat{\sigma}_{k}^{12} + {\rm h.c.} \right]$ describes the coherent energy exchange between the atomic transition $1_k\to 2_k$ and the optical cavity with the strength $
g_{21} = - (g_{31}\Omega)/ (\omega_{32} - \omega_d - i\gamma_{31}/2)$. The emerging Lindblad term in Eq.~\eqref{eq:eff-qme} describes the Raman transition-induced decay with the rate $\gamma_{21}=\gamma_{31}\Omega^2/|\omega_{32}-\omega_d-i\gamma_{31}/2|^2$. Eq.~\eqref{eq:eff-qme} has a similar form as the quantum master equation for the superradiant laser~\citep{DMeiser2009}, except that  $g_{21}$ and $\gamma_{21}$ can be actively controlled by adjusting the parameters $\Omega,\omega_d$ of the dressed laser.

To simulate systems with tens of thousands of atoms, we solve the quantum master equation~\eqref{eq:qme} and~\eqref{eq:eff-qme} with the cumulant mean-field approach. In this approach, we derive the equations $\partial_t \left\langle \hat{o} \right\rangle = {\rm tr}\{(\partial_t \hat{\rho} ) \hat{o}\}$ for the mean values $\left\langle \hat{o} \right\rangle$ (for any operator $\hat{o}$), and obtain a hierarchy of inter-dependent equations for the mean values of many operators due to the interaction or the collective process, and apply the cumulant expansion approximation,  e.g. $\left \langle \hat{o}\hat{p}\hat{q} \right \rangle\approx\left \langle \hat{o}\right \rangle \left \langle \hat{p}\hat{q} \right \rangle + \left \langle \hat{p}\right \rangle \left \langle \hat{o}\hat{q} \right \rangle + \left \langle \hat{q}\right \rangle \left \langle \hat{o}\hat{p} \right \rangle - 2 \left \langle \hat{o}\right \rangle \left \langle \hat{p} \right \rangle \left \langle \hat{q} \right \rangle$ (for any operators $\hat{o},\hat{p},\hat{q}$), to remove the hierarchy and obtain a closed set of equations. If all the atoms are  identical  the terms $\left \langle \hat{\sigma}^{lm}_k \right \rangle$, $\left \langle \hat{a}^\dagger \hat{\sigma}_k^{23} \right \rangle$ are the same for all the atoms, and the terms $\left \langle \hat{\sigma}_k^{lm} \hat{\sigma}_{k'}^{l'm'} \right \rangle$ are same for all atom pairs $(k,k')$. In this way, we can reduce the number of coupled equations from the order of $\sim N^3 $ to a few tens. In practice, we employ the QuantumCumulant.jl package~\citep{DPlankensteiner} to implement the above procedure, and summarize the corresponding codes in Appendix \ref{sec:codes}. To calculate the superradiant Raman scattering spectrum, we have also reformulated the codes to derive the equations for the two-time correlation functions, and ensured that they work also in  the interaction picture as employed here.

We employ the Dicke state picture to analyze the collective dynamics of the atomic ensemble. To this end, we define firstly the collective operators $\hat{J}_{x(y)}=\frac{1(i)}{2}\sum_{k=1}^{N}\left(\hat{\sigma}_{k}^{12}\pm\hat{\sigma}_{k}^{21}\right)$,$\hat{J}_{z}=\frac{1}{2}\sum_{k=1}^{N}\left(2\hat{\sigma}_{k}^{22}-\hat{1}_{k}\right)$, and their square   $\hat{J}_{x\left(y\right)}^{2}=\frac{1}{4}\sum_{k}^{N}\hat{1}_{k}+\frac{1}{4}\sum_{k\neq k'}^{N}(\hat{\sigma}_{k}^{12}\hat{\sigma}_{k'}^{12}\pm\hat{\sigma}_{k}^{12}\hat{\sigma}_{k'}^{21}\pm\hat{\sigma}_{k}^{21}\hat{\sigma}_{k'}^{12}+\hat{\sigma}_{k}^{21}\hat{\sigma}_{k'}^{21})$,
and $\hat{J}_{z}^{2}=\frac{1}{4}\sum_{k}^{N}\hat{1}_{k}+\frac{1}{4}\sum_{k\neq k'}^{N}(4\hat{\sigma}_{k}^{22}\hat{\sigma}_{k'}^{22}-2\hat{\sigma}_{k}^{22}\hat{1}_{k'}-2\hat{1}_{k}\hat{\sigma}_{k'}^{22}+\hat{1}_{k}\hat{1}_{k'})$.
Here, $\hat{1}_{k}$ are the identity operators. Then, we define the Dicke states as the eigen states of the
equations $\left(\sum_{i=x,yz}\hat{J}_{i}^{2}\right)\left|J,M\right\rangle =J\left(J+1\right)\left|J,M\right\rangle $,
$\hat{J}_{z}\left|J,M\right\rangle =M\left|J,M\right\rangle $, where
the integer or half-integer $J=0,...,N/2$
indicates the degree of symmetry (the coupling strength) and the number $M$ in the range $-J\le M\le J$ labels the degree of the excitation.  Inspired by the above equations, we introduce the mean values $\bar{J},\bar{M}$ through the equations $\bar{J}\left(\bar{J}+1\right)=\sum_{i=x,yz}\left\langle \hat{J}_{i}^{2}\right\rangle $
and $\bar{M}=\left\langle \hat{J}_{z}\right\rangle $. Note that all the processes of two-level systems, as described by the effective quantum master equation, have been identified in the Dicke state picture ~\citep{ShammahN,ZhangY2018}. If all the atoms  are identical, the expectation values of the collective operators can be calculated from the expressions $\left\langle \hat{J}_{x\left(y\right)}^{2}\right\rangle =\frac{1}{4}N\pm\frac{1}{4}N\left(N-1\right)(\left\langle \hat{\sigma}_{1}^{12}\hat{\sigma}_{2}^{12}\right\rangle \pm\left\langle \hat{\sigma}_{1}^{12}\hat{\sigma}_{2}^{21}\right\rangle \pm\left\langle \hat{\sigma}_{1}^{21}\hat{\sigma}_{2}^{12}\right\rangle +\left\langle \hat{\sigma}_{1}^{21}\hat{\sigma}_{2}^{21}\right\rangle )$,
and $\left\langle \hat{J}_{z}^{2}\right\rangle =\frac{1}{4}N+\frac{1}{4}N\left(N-1\right)\left(4\left\langle \hat{\sigma}_{1}^{22}\hat{\sigma}_{2}^{22}\right\rangle -4\left\langle \hat{\sigma}_{1}^{22}\right\rangle +1\right)$,
as well as $\left\langle \hat{J}_{x}\right\rangle =\frac{1}{2}N(\left\langle \hat{\sigma}_{1}^{12}\right\rangle +\left\langle \hat{\sigma}_{1}^{21}\right\rangle ),\left\langle \hat{J}_{y}\right\rangle =\frac{i}{2}N(\left\langle \hat{\sigma}_{1}^{12}\right\rangle -\left\langle \hat{\sigma}_{1}^{21}\right\rangle),\left\langle \hat{J}_{z}\right\rangle =\frac{1}{2}N(2\left\langle \hat{\sigma}_{1}^{22}\right\rangle -1)$.  As a complementary
picture of the collective process, we can also calculate the collective spin vector (also known as Bloch vector), $\mathbf{A}=\sum_{i=x,y,z}\left\langle \hat{J}_{i}\right\rangle \mathbf{e}_{i}$
with the expectation values $\left\langle \hat{J}_{i}\right\rangle =\sum_{\alpha=1,2}\left\langle \hat{J}_{i}\right\rangle $
and the unit vectors $\mathbf{e}_{i}$ of Cartesian coordinate system. However, our calculations show that the vector components $A_x,A_y$ are always zero within the pulsed and continuous superradiant Raman scattering, and thus the collective spin vector is not a good picture to analyze the cause of these scattering phenomena.

Before presenting the numerical results, we comment shortly on the used parameters. The optical cavity has a frequency $\omega_c=2\pi \times(3.77\times{10}^{14}+6.8\times{10}^9)$ Hz and a 
damping rate $\kappa=2\pi\times11$ MHz. The atoms couple with the cavity with a strength  $g_{31} = 2\pi\times 506$ kHz, and have the transition frequencies  $\omega_{32} =2\pi\times(3.77\times{10}^{14}-2\times{10}^9)$ GHz,  $\omega_{21} =2\pi\times6.8$ GHz, and the decay rate   $\gamma_{31} = 2\pi\times5.75$ MHz.  The dressed laser with a frequency $\omega_d =\omega_{32}+2\pi\times2$ GHz couples with the atoms with a strength $\Omega=2\pi\times 5$ MHz.

\begin{figure}[!htp]
\centering
\includegraphics[scale=0.23]{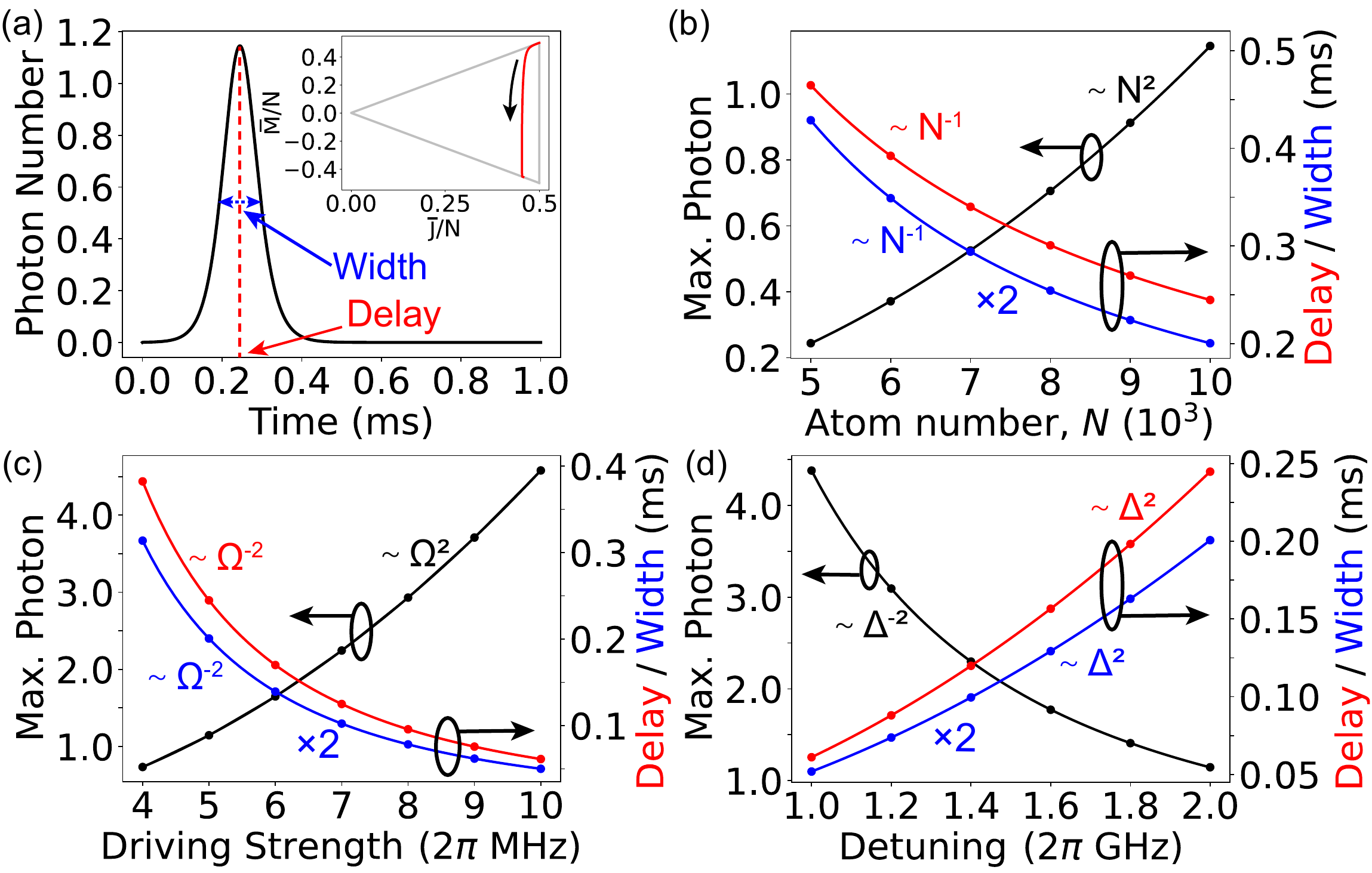}
\caption{\label{fig:pulses-weak}Superradiant Raman scattering pulses for systems within the crossover coupling regime. Panel (a) shows the dynamics of the intra-cavity photon number, and of the atomic ensemble in the Dicke states space (inset).  Here, the average Dicke numbers are defined with respect to the two hyper-fine ground levels. Panel (b) shows the evolution of the maximum, width and delay time of the pulses as function of the number of atoms. Panel (c,d) show the similar results as in the panel (b) except that the strength $\Omega$ and frequency detuning $\Delta=\omega_l-\omega_{32}$ of the dressed laser are varied.  Here, the atoms are assumed to be initially on the higher hyper-fine ground level, and the key reference parameters are $N =10^4$, $\Omega=2\pi\times5$ MHz, and $\omega_d = \omega_{32} + 2\pi\times2 {\rm  GHz}$. }
\end{figure}

\section{Superradiant Raman Scattering Pulses \label{sec:pulses}}

\begin{figure}[!htp]
\centering
\includegraphics[scale=0.23]{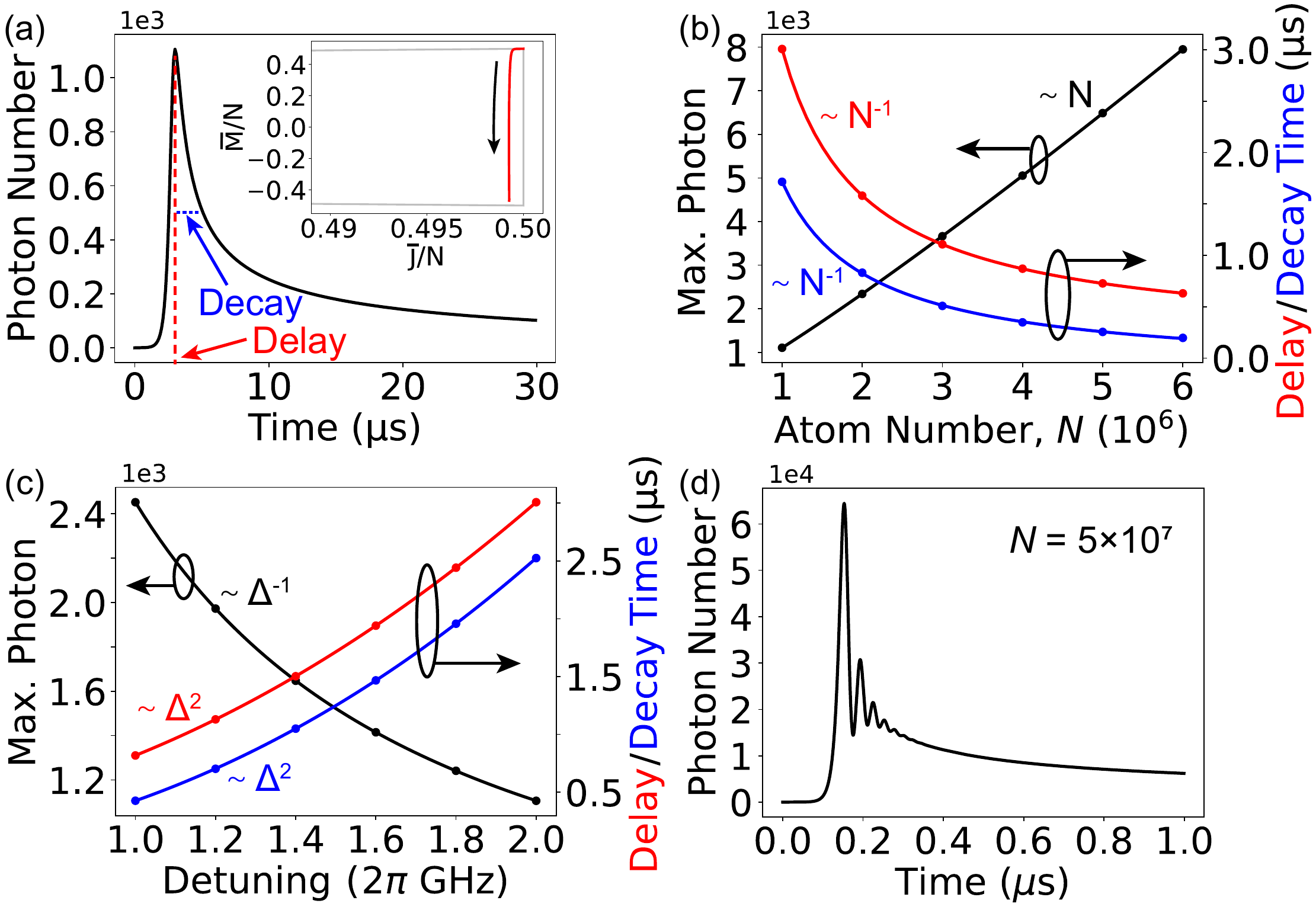}
\caption{\label{fig:pulses-strong} Superradiant Raman scattering pulses for systems within the strong coupling regime.  Panel (a-c) show the similar results as in Fig. 2(a,b,d) except that the decay time of the tail is considered instead of the pulse width. Here, the decay time is defined as the time to reach the half of the maximum value. Panel (d) shows the photon number for the systems with $5\times 10^7$ atoms. The key reference parameters are $N =10^6$, $\Omega=2\pi\times5$ MHz, and $\omega_d = \omega_{32} + 2\pi\times2 {\rm  GHz}$. }
\end{figure}

In the following, we study the superradiant Raman scattering pulses, and find the different behaviors for the systems in the crossover regime (Fig.~\ref{fig:pulses-weak}) and the strong coupling regime (Fig.~\ref{fig:pulses-strong}). In the former regime, the collective coupling $\sqrt{N}g_{31}$ is slightly larger than the photon loss rate $\kappa$, and the dressed laser drives directly the bare atoms.  In the latter regime, $\sqrt{N}g_{31}$ is significantly larger than $\kappa$, and the dressed laser derives the atoms-cavity system through the formed atom-photon dressed states. Here, we do not discuss the response for the system in the weak coupling regime, since it is similar to that for the system in the crossover regime. 

We firstly focus on the systems in the crossover regime. We find that  once the dressing laser excites the atoms initially on the higher hyper-fine ground level, the intra-cavity photon number increases firstly and then decreases, forming a pulse [Fig.~\ref{fig:pulses-weak}(a)]. Accompanying with this evolution, the population transfers from the higher to the lower hyper-fine ground level (not shown). Since the excited level is not populated at all, this is clearly a Raman process. As a complement to the population change, the ensemble occupies initially the top-right corner of the Dicke states space, and declines vertically with a finite distance from the right boundary to the lower boundary [inset of Fig.~\ref{fig:pulses-weak}(a)], which indicates the superradiant origin of the observed pulses. To prove  this point further, we examine also the maximum, width and delay time of the pulses as function of number of atoms [Fig.~\ref{fig:pulses-weak}(b)]. We find that the maximum increases quadratically, the width and delay time decrease inversely, which are the characteristics of the superradiant pulses~\citep{MANorcia2016}. 
The same results are also achieved with the effective quantum master equation (not shown). 

The above results verify that the observed phenomena are superradiant pulses. However, in contrast to the systems with the true atomic transition, here, the collective coupling with the cavity mode is mediated by Raman scattering, and can be controlled by the dressed laser. To demonstrate this point, we study further how the dressed laser parameters $\Omega,\omega_d$ affect the superradiant pulses [Fig.~\ref{fig:pulses-weak}(c,d)]. We find that as the driving strength $\Omega$ increases, the pulse maximum  increases quadratically ($\sim\Omega^2$), while the pulse width and delay time decrease quadratically ($\sim\Omega^{-2}$). On the contrary, as the detuning $\omega_d-\omega_{32}$ increases, the pulse maximum decreases quadratically, while the pulse width and delay time increase quadratically. All these can be understood by examining effective master equation~\eqref{eq:eff-qme}. In this equation, the effective coupling strength $g_{21}$ is proportional roughly to the ratio $\Omega/(\omega_d-\omega_{32})$, and the effective Purcell-enhanced decay rate  $\Gamma \approx 4g_{21}^2/\kappa$ scales quadratically with this ratio. Although the effective decay rate $\gamma_{21}$ scales also quadratically with this ratio, it is relatively small than the collective decay rates $(J+M)(J-M+1)\Gamma$ of the Dicke states $\left |J,M\right \rangle$ with $J\approx N/2$.

We then consider the systems in the strong coupling regime. In this case, after the switch-on of the dressed laser, the intra-cavity photon number increases dramatically in short time, but decays slowly in long time, resulting to a distorted pulse [Fig.~\ref{fig:pulses-strong}(a)]. Following this dynamics, the population transfer from the higher to lower hyper-fine ground level is fast in short time, but becomes slow in long time (not shown). Here, the atomic ensemble decays also vertically, but the evolution is much closer to the right boundary [inset of Fig.~\ref{fig:pulses-strong}(a)].  The further analysis indicates  the pulse maximum scales linearly with number of atoms, the pulse delay time and the slow decay time of the pulse decrease inversely [Fig.~\ref{fig:pulses-strong}(b)]. With the effective master equation~\eqref{eq:eff-qme}, we obtain the similar results as before (Fig.~\ref{fig:eff} in the Appendix). Thus, the observed distorted pulse goes beyond the scope of the effective model, and originates from the full model. 

We study further the dependence of the distorted pulses on the dressed laser parameters $\Omega,\omega_d$. The pulse maximum, delay and decay time behave similarly for different dressed strength $\Omega$ (Fig.~\ref{fig:pulses-strong2} of the Appendix) and frequency detunning $\omega_d-\omega_{32}$  as before  except that  the pulse maximum scales inversely with the frequency detunning [Fig.~\ref{fig:pulses-strong}(c)]. For the systems with more atoms, we find that the decay at the longer time evolves into the decayed oscillations [Fig.~\ref{fig:pulses-strong}(d)]. Since this oscillation resembles the Rabi oscillations in the strong coupling regime~\citep{MANorcia2016-1}, we attribute the distorted pulses to the behavior in the strong coupling regime.  We note that the distorted pulses agree qualitatively with Fig. 1(c) of the experimental paper~\citep{JDBohnet2012}, while the quadratic dependence of the pulse maximum agrees qualitatively with Fig. 1(e) of that paper. This comparison indicates that these results in the experiment might be caused by the systems in the different regimes.

\begin{figure}[!htp]
\centering
\includegraphics[scale=0.23]{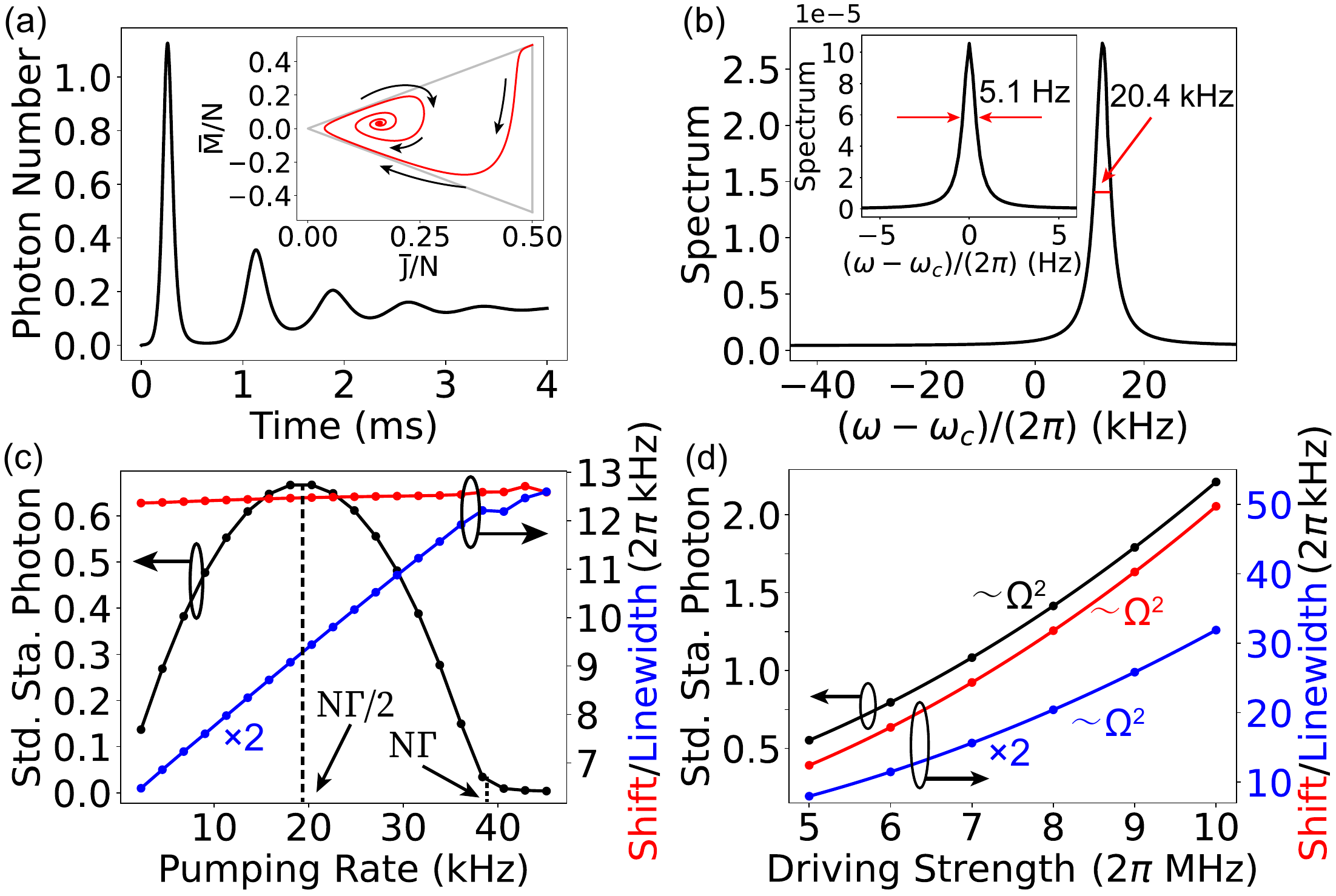}
\caption{\label{fig:continuous-weak} Continuous superradiant Raman scattering for systems within the crossover coupling regime. Panel (a)  shows the evolution of the intra-cavity photon number and the atomic ensemble in the Dicke state space (inset). Panel (b) compares the shifted and broad Raman scattering spectrum computed  with the full model with the centered and narrower spectrum calculated with the effective model (inset).  Panel (c,d) show the steady-state intra-cavity photon number (left axes), the spectral frequency shift and linewidth (right axes) as function of the incoherent pumping rate and the strength of the dressed laser, respectively.  In the panel (c), the vertical lines show $N\Gamma/2$ and $N\Gamma$ with the Purcell-enhanced decay rate $\Gamma\approx 4 g_{12}^2/\kappa$. Here, we assume that the atoms are incoherently pumped from the lower to higher hyper-fine ground level. }
\end{figure}

\section{Continuous Superradiant Raman Scattering \label{sec:continuous}}

In the following, we consider the systems in the presence of the incoherent atomic pumping. Similar to the above discussion, here, we also distinguish the systems in the crossover (Fig.~\ref{fig:continuous-weak}) and strong coupling regime (Fig.~\ref{fig:continuous-strong}), and focus on the former case firstly. In the presence of the pumping, the intra-cavity photon number demonstrates also several peaks after the first main peak, and  evolves eventually to a constant value about 0.135 [Fig.~\ref{fig:continuous-weak}(a)]. In accordance with this, the atomic ensemble decays vertically after climbing the lower and upper boundary, and reaches eventually to a point slightly above the left-most corner after repeating the similar dynamics [inset of Fig.~\ref{fig:continuous-weak}(a)]. When the atomic ensemble reaches the steady-state, we can calculate the corresponding Raman scattering spectrum  [Fig.~\ref{fig:continuous-weak}(b)]. The spectrum shows a peak with a frequency about $2\pi\times12.4$ kHz shifted from the frequency of the optical cavity mode, and a linewidth of $2\pi\times3.24$ kHz. If we consider the effective model, we obtain the similar results as Fig.~\ref{fig:continuous-weak}(a) for the dynamics, but a sharp spectrum with a frequency around the cavity mode, and a linewidth of $2\pi \times 0.82$ Hz. The observed shifted and broad peak agrees qualitatively with Fig. 4(a) of the experimental paper~\citep{JDBohnet2012}. 


We have further studied the intra-cavity photon number, the spectral shift and linewidth as function of the incoherent pumping rate [Fig.~\ref{fig:continuous-weak}(c)]. As the pumping rate $\gamma_{12}$ increases, the photon number increases gradually, and approaches a maximum for $\gamma_{12}= N\Gamma/2$, and then drops dramatically to a value close to zero for $\gamma_{12}= N\Gamma$. Here, $\Gamma\approx 4 g_{12}^2/\kappa$ is the effective Purcell-enhanced decay rate. This result agrees qualitatively with Fig. 2 of the experimental article~\citep{JDBohnet2012}. Accompanying with this, the spectral shift increases linearly within one kilohertz, while the linewidth increases linearly from $2\pi \times 3$ kHz to $2\pi \times 6$ kHz. In contrast, if the effective model is adopted, the photon number behaves similarly but the linewidth behaves in a opposite way, and achieves a minimal value of $2\pi\times 0.62$ Hz (Fig.~\ref{fig:eff} of the Appendix). This comparison indicates again the difference of the full and effective model. To reveal the physics leading to the difference, we further study the influence of the driving strength of the dressed laser [Fig.~\ref{fig:continuous-weak}(d)]. We find that the photon number, the frequency shift, and the linewidth all increase quadratically with the strength. To explain the cause of the frequency shift, we note that  the atom will experience a frequency shift $\Delta\omega = \Omega^2/4\Delta$, i.e. an AC Stark shift, if it is driven by a laser with a strength $\Omega$ and a frequency detuning $\Delta$. The $\Omega^2$-dependence of such a shift coincides with what shown in Fig.~\ref{fig:continuous-weak}(d), and 
the $1/\Delta$-dependence is further illustrated in Fig. ~\ref{fig:detunung shift}. Thus, it is highly possible that the frequency shift as observed here is caused by AC Stark shift.

In Fig.~\ref{fig:continuous-strong} of the Appendix, we have studied further the continuous superradiant Raman scattering for systems within the strong coupling regime and in the presence of incoherent atomic pumping. These results agree qualitatively with the those for the system within the crossover regime, except that the frequency shift is orders of magnitude smaller and the linewidth broadening is also much smaller.

\section{Conclusion}

In summary, to understand the experiment~\citep{JDBohnet2012} of J. G. Bohnet et al.  proving the concept of the superradiant laser~\citep{DMeiser2009}, we have developed a quantum master equation to describe the superradiant Raman scattering of three-level atoms coupled with an optical cavity and a dressed laser, and obtained an effective quantum master equations for the system with two-level atoms by adiabatically eliminating the higher excited level. We have solved these equations with cumulant mean-field approach to study the dynamics of the systems with many atoms, and modified codes in the QuantumCumulant.jl package to calculate the Raman scattering spectrum. 

Through the numerical simulations, we distinguish the normal and distorted superradiant Raman pulses for the systems within the crossover and strong coupling regime in the absence of incoherent pumping, which agree qualitatively with the experimental results and provide a unified view on these results. More importantly, our calculations demonstrate the shifted and broad spectrum for the continuous Raman scattering in the presence of the incoherent pumping, which agrees also qualitatively with the experimental results and provides also an alternative explanation on the observed broad spectrum rather the expected narrow spectrum. In any case, our study points out the rich physics involved in the superradiant Raman scattering, and further study could explore the application of this phenomenon in the magnetic field sensing~\citep{JMWeiner2012}, the real-time track of quantum phase~\citep{AShankar2019}, the Dicke phase transition of non-equilibrium dynamics~\citep{MPBaden2019}, and so on.

\begin{acknowledgments}
Huihui Yu carried out the numerical calculations under the supervision of Yuan Zhang who developed the theory and the numerical programs. They contribute equally to the work. All authors contributed to the analyses and the writing of the manuscript. This work is supported by the National Key R\&D Program of China under Grant No. 2021YFA1400900, the National Natural Science Foundation of China under Grants No. 12004344, 12174347, 12074232, 12125406, 62027816, U21A2070, and the Cross-disciplinary Innovative Research Group Project of Henan Province No. 232300421004, as well as the Danish National Research Foundation through the Center of Excellence for Complex Quantum Systems (Grant agreement No. DNRF156). 
\end{acknowledgments}

\newpage
\appendix
\renewcommand\thefigure{A\arabic{figure}}
\renewcommand\thetable{A\arabic{table}}
\renewcommand{\bibnumfmt}[1]{[S#1]}
\setcounter{figure}{0}

\section{\label{sec:codes}Julia Codes}

\begin{figure}
\begin{centering}
\includegraphics[scale=0.47]{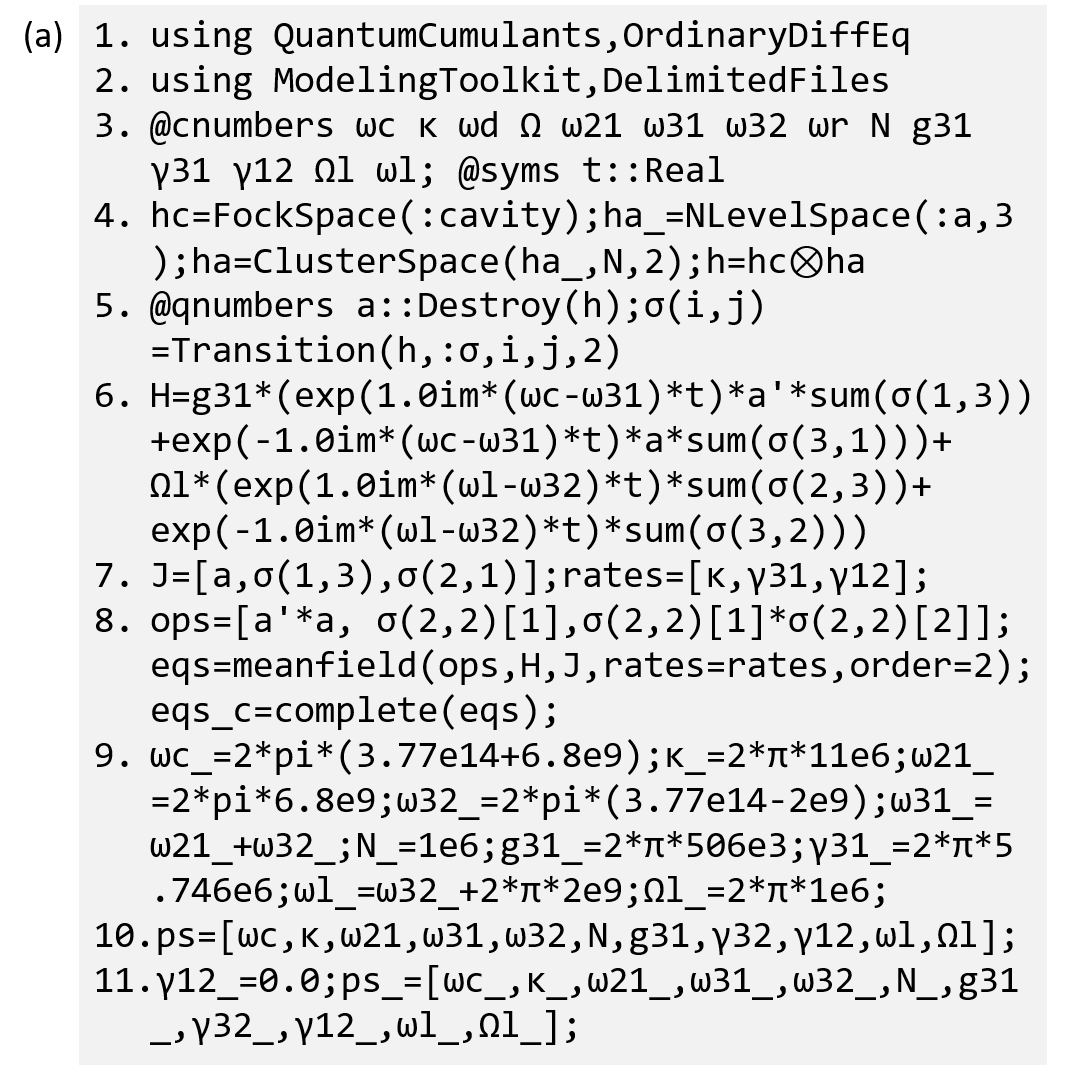} 
\includegraphics[scale=0.685]{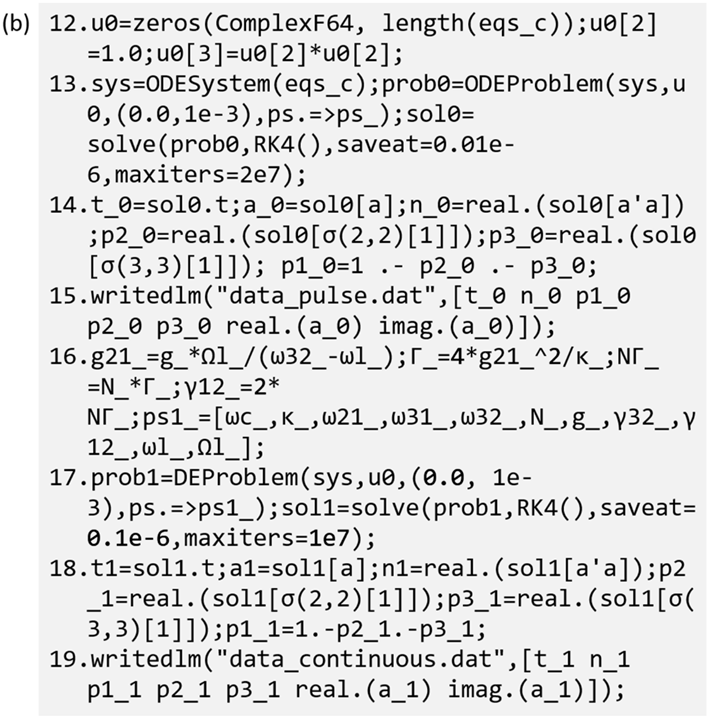} 
\par\end{centering}
\caption{\label{fig:codes}Julia codes to solve the quantum master equation~\eqref{eq:qme} with the mean-field approach. The details of the codes are explained in the text. }
\end{figure}

In this Appendix, we explain the Julia codes to solve the quantum master equations. First, we focus on the system with three-level atoms, and present the codes to solve Eq.~\eqref{eq:qme} (Fig.~\ref{fig:codes}). The 1st and 2nd line import the necessary packages. The 3rd line defines the complex numbers and the time argument. The 4th line defines the Hilbert space for the cavity, the single three-level atom, and the atomic ensemble, as well the total system. The 5th line defines the photon annihilation operator, the transition and projection operators of the atoms. The 6th line defines the system Hamiltonian in the interaction picture. The 7th line defines the list of operators and rates, which are used later on to specify the Lindblad dissipative superoperators. The 8th line defines a list of three operators, and derive the equations for the mean values of these operators, as well as obtain the closed set of mean-field equations by applying second-order cumulant expansion approximation. The 9th line specifies the related parameters. The 10th line defines a list of the parameters, and the 11th line specifies a list of their values. Note that for the superradiant Raman scattering pulses, the incoherent pumping rate $\gamma_{12}$ is assumed as zero. 

The 12th line specifies the initial values of the mean-fields. Here, we assume that the atoms are initially on the excited state. The 13th line defines the ordinary differential equation (ODE) system with the derived equations, the ODE problem with the initial values, the time range and the parameters, and then solve the ODE equations with Runge-Kutta method. The 14th line extracts the time list, the intra-cavity field amplitude, the intra-cavity photon number, the populations on the two hyper-fine ground levels and the excited level. The 15th line saves the data into a text file. The 16th to 19th line repeat the same procedure as the 9th to 15th line, except that the incoherent pumping rate $\gamma_{12}$ is not zero, and the continuous superradiant Raman scattering is obtained.

\begin{figure}
\begin{centering}
\includegraphics[scale=0.41]{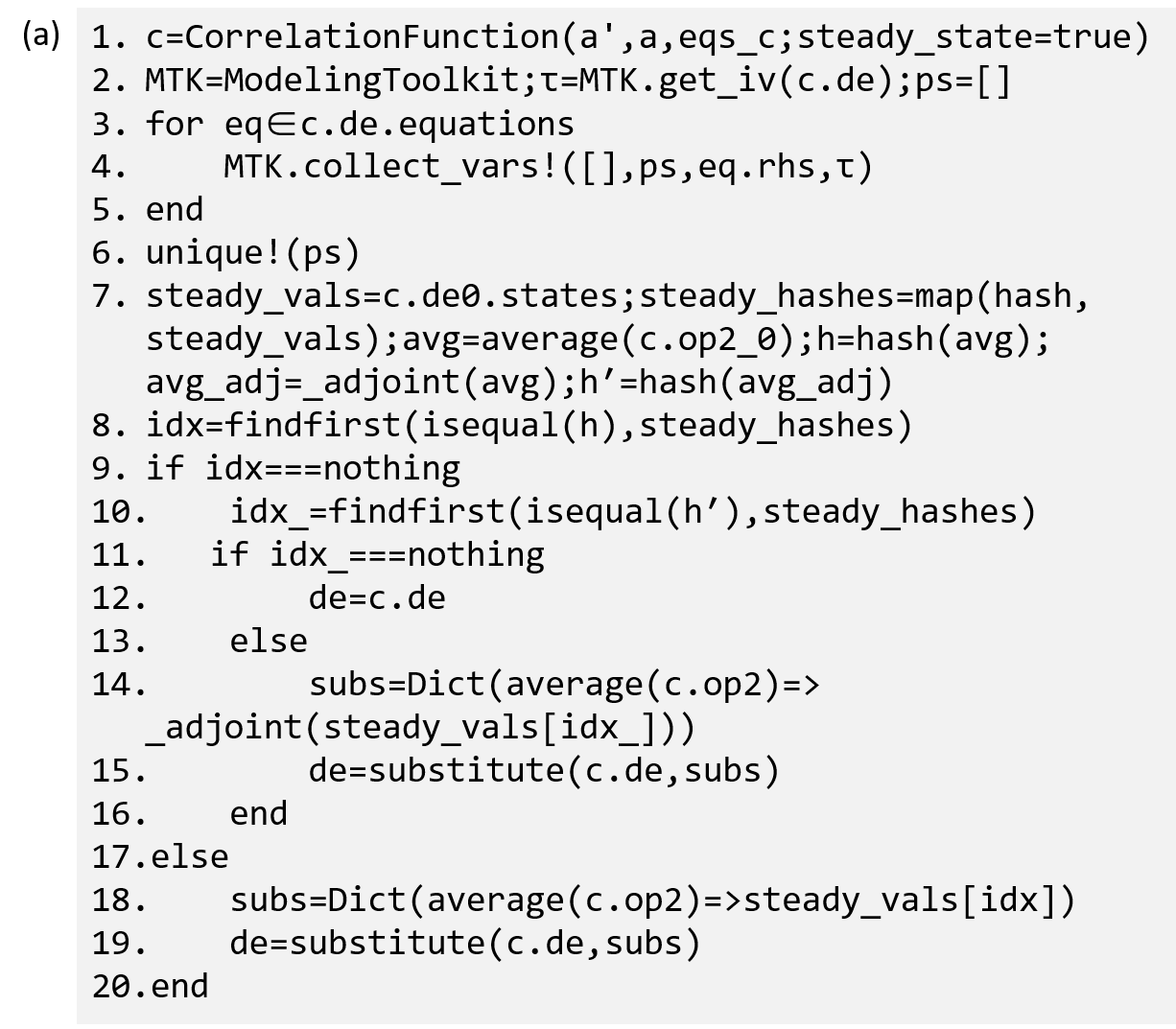}
\includegraphics[scale=0.39]{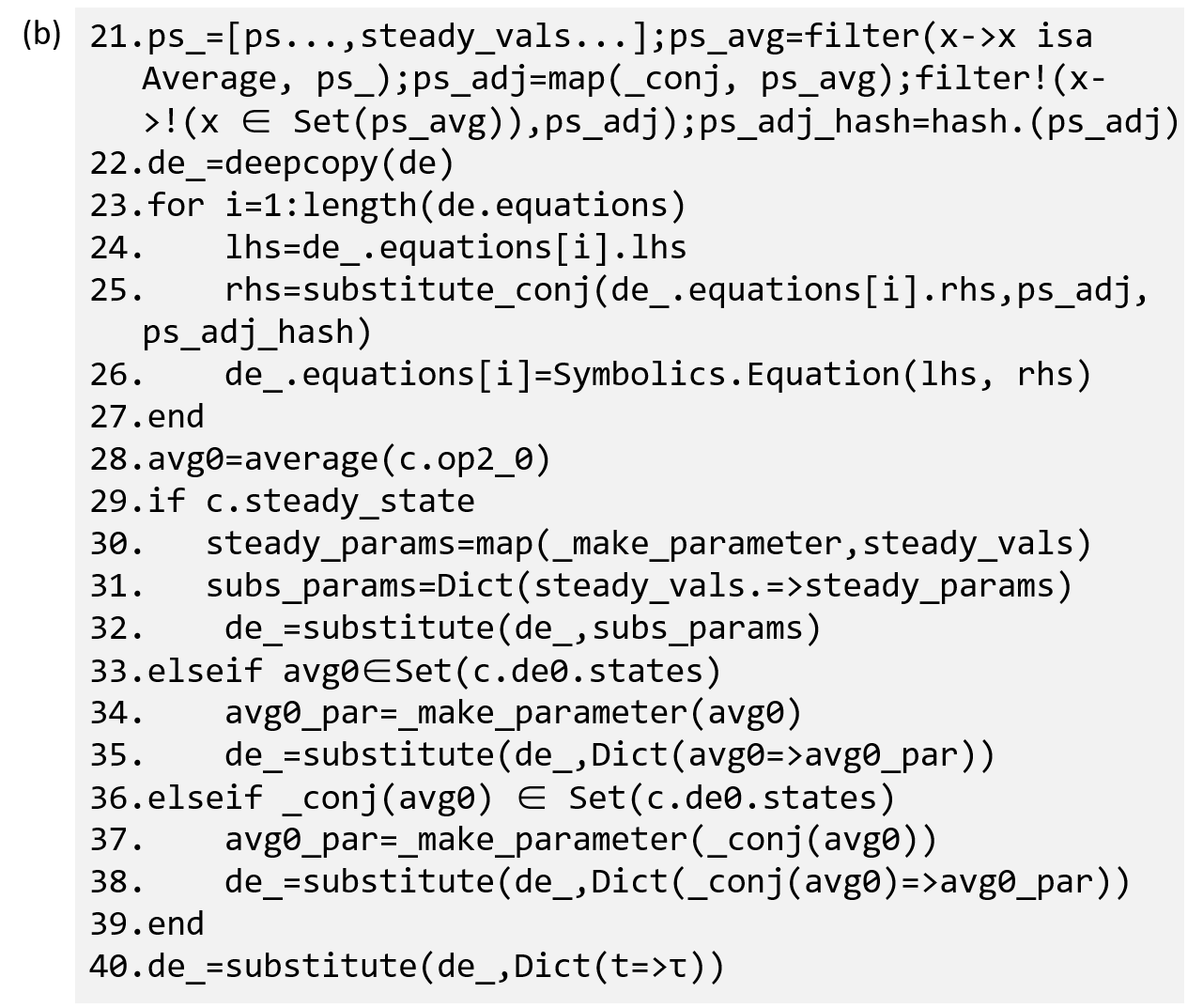} 
\includegraphics[scale=0.415]{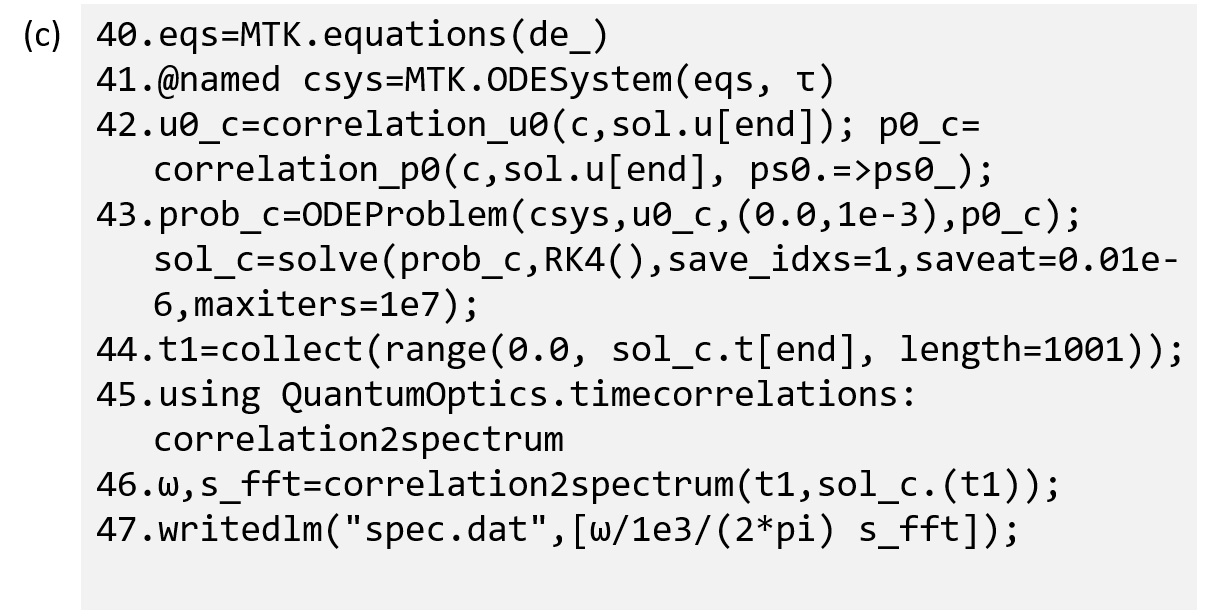} 
\par\end{centering}
\caption{\label{fig:spectrum}Julia codes to calculate the superradiant Raman scattering spectrum for the system at the steady-state. }
\end{figure}

The QuantumCumulant.jl package has provided convenient functions to derive the equations for the two-time correlation function, and to calculate the system spectrum. However, this works only in the case with the time-independent Hamiltonian. In order to compute the superradiant Raman scattering, in the current study, we have reformulated the master equation in an interaction picture, where the Hamiltonian becomes time-dependent. As a result, we need to reformulate the aforementioned codes (Fig.~\ref{fig:spectrum}). 

In Fig.~\ref{fig:spectrum}(a), the 1st line derives the equations for the two-time correlation functions $\left \langle \hat{o}_1(\tau) \hat{o}_2(0)\right \rangle $. Unfortunately, the derived equations can not been directly passed to the functions in the ModelingToolkit.jl package to define the mathematical model and the ODE system. The 2nd line defines the abbreviation for this package,   the time argument, and the empty list for the variables. The 3rd to 5th line analyze  the derived equations, and collect the variables. The 6th line analyzes the list of variables, and removes the abundance. The 7th line evaluates the mean-values at the steady-state, which are used later on to calculate the initial conditions for the two-time correlation function. The 8th to 20th line replace $\left \langle \hat{o}_2(0) \right \rangle$ with the steady-state variables. The 21th line defines the list of variables, which includes also the steady-state variables. The 22th line copies all the derived equations. The 23th to 39th line construct the object to the two-time correlation functions. The 40th line defines the mathematical model with the derived equations, and the 41th defines the ODE system. The 42th line specifies the initial values of the correlation functions,  the list of values for the parameters. The 43th line defines the ODE problem and solve it with Runge-Kutta method. The 44th line defines a list of time point. The 45th line imports the "correlation2spectrum" function from the QuantumOptics.jl package. The 46th line carries out the Fourier transform to the correlation functions, and returns the frequency and the spectrum. The 47th line saves the data into a text file. 

\begin{figure}
\begin{centering}
\includegraphics[scale=0.38]{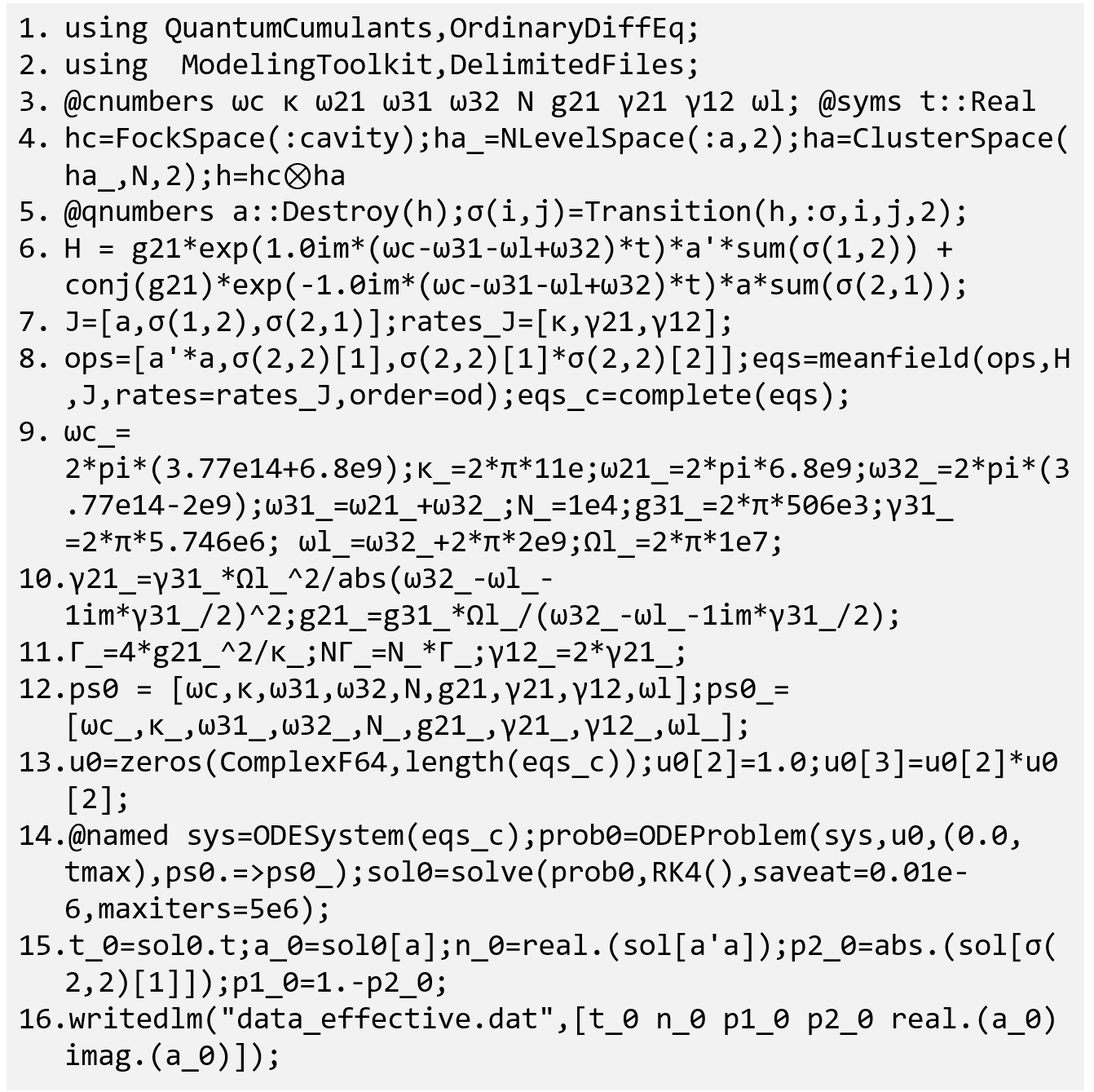}
\par\end{centering}
\caption{\label{fig:codes-eff}Julia codes to solve the effective quantum master equation~\eqref{eq:eff-qme}. }
\end{figure}

Furthermore, we present the codes to solve the effective quantum master equation~\eqref{eq:eff-qme} (Fig.~\ref{fig:codes-eff}). The most codes are similar as those in Fig.~\ref{fig:codes} except for several small differences. In the 4th line, we define the atoms as two-level systems. In the 6th line, we defines effective Hamiltnoian in the interaction picture. In the 7th line, we account for the effective decay process in the Lindblad term. The 10th line, we calculate the effective coupling strength $g_{12}$ and the effective decay rate $\gamma_{21}$. In the 15th line, we extract only the population of the two hyper-fine ground levels.

\section{Extra Numerical Results}

In this Appendix, we provide extra results to complement those in the main text. 

\begin{figure}[!htp]
\centering
\includegraphics[scale=0.23]{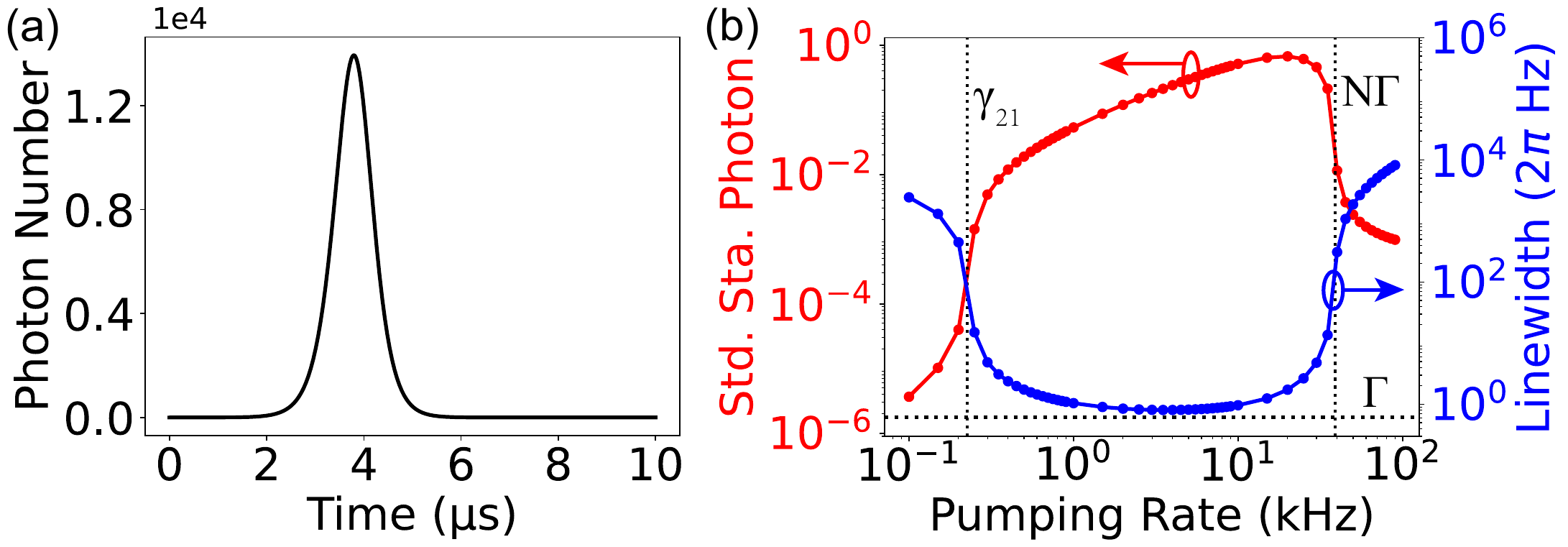}
\caption{\label{fig:eff} Simulations with the effective model. Panel (a) shows the evolution of the intra-cavity photon number for the system with $10^6$ atoms within the strong coupling regime. Panel (b) shows the steady-state intra-activity photon number (left axis) and the spectral linewidth (right axis) as function of the incoherent pumping rate for the system within the crossover regime. The two vertical dashed lines represent the effective decay rate $\gamma_{21}$ (left) and the collective enhanced decay rate $N\Gamma$ (right), while the vertical line represents the Purcell-enhanced decay rate $\Gamma$. }
\end{figure}

\subsection{Calculations with Effective Model}

Figure~\ref{fig:eff} shows the simulations calculated with the effective model. For the system with $10^6$ atoms initially prepared on the upper hyper-fine ground level in the absence of incoherent pumping, the simulation shows a superradiant pulse [Fig.~\ref{fig:eff}(a)] just like the system within the weak coupling regime, which contradicts with the simulations calculated with the full model [Fig.~\eqref{fig:pulses-strong}(a)] and with the experimental results [Fig. 1(c) of Ref.~\citep{JDBohnet2012}]. 

For the system with $10^4$ atoms in the presence of the incoherent pumping, the intra-cavity photon number increases firstly and then decreases, which agrees with what calculated with the full model  [Fig.~\ref{fig:continuous-weak}(c), left axis of  Fig.~\ref{fig:eff}(b)] and the experimental result [Fig. 2(a) of Ref. ~\citep{JDBohnet2012}]. At the same time, the spectral linewidth behaves oppositely, and approaches the minimal value around hertz for moderate pumping [right axis of Fig.~\ref{fig:eff}(b)], which contradicts with what obtained with the full model [Fig.~\ref{fig:continuous-weak}(c)]. Although  the  small linewidth agrees with what expected, it seems not agree with the experimental result [Fig. 4(a) of Ref. ~\citep{JDBohnet2012}].

\begin{figure}[!htp]
\centering
\includegraphics[scale=0.32]{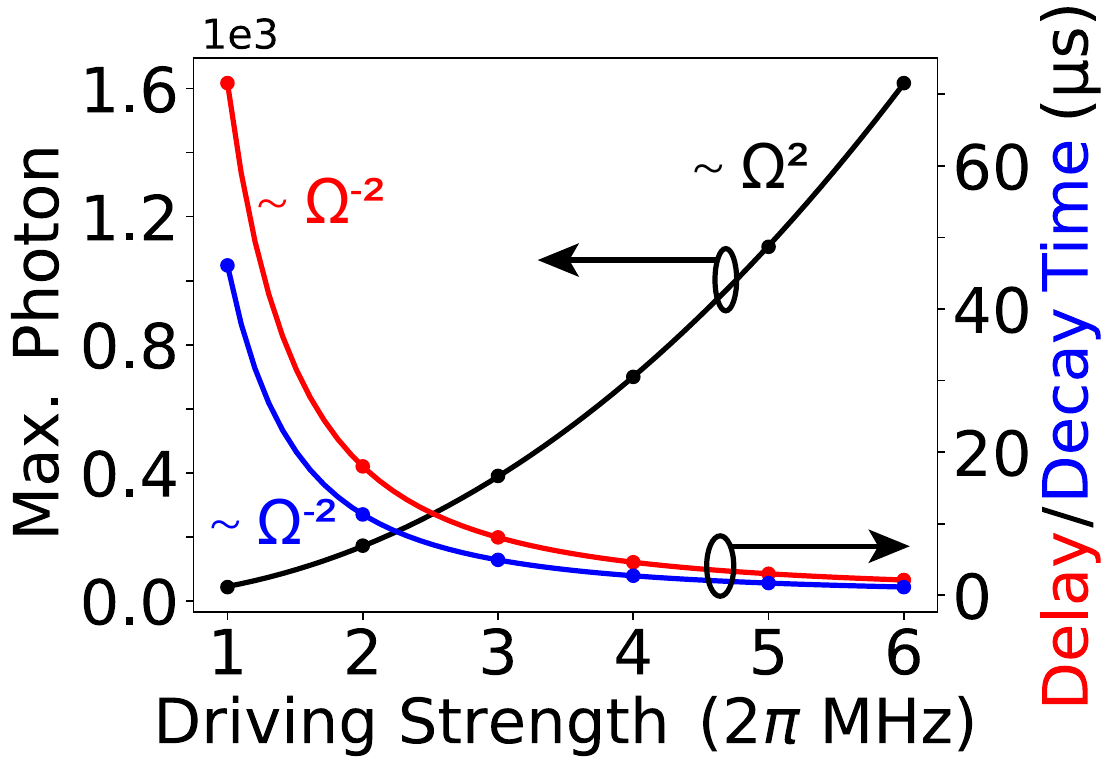}
\caption{\label{fig:pulses-strong2} Evolution of the maximum, delay and decay time of the pulses as function of the dressed strength. }
\end{figure}

\subsection{Extra Results for Superradiant Raman Scattering Pulses within the Strong Coupling Regime}

Figure~\ref{fig:pulses-strong} demonstrates how the maximum photon number, delay and decay time of the superradiant Raman pulses change with number of atoms and frequency detuning within the strong coupling regime. In Fig.~\ref{fig:pulses-strong2}, we complement these   by showing the results as function of  the strength $\Omega$ of the driving laser. We find that the maximal photon number scales as $~\Omega^2$ while the pulse decay and decay time scale as  $~\Omega^{-2}$. These results agree with those in the crossover regime (Fig.~\ref{fig:pulses-weak}c).

\begin{figure}[!htp]
\centering
\includegraphics[scale=0.32 ]{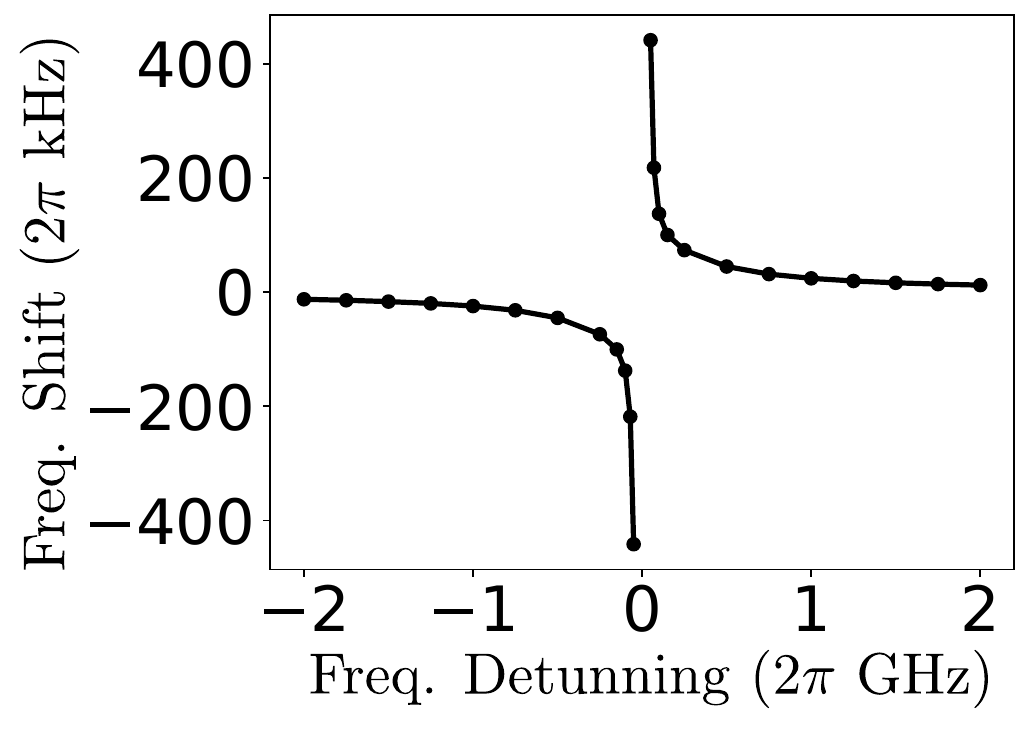}
\caption{\label{fig:detunung shift} Frequency shift (relative to the cavity mode) in the steady-state superradiant Raman spectrum for different frequency detunning of the dressed laser to the atomic transition between the upper hyper-fine ground state and the excited state. } 
\end{figure}

\subsection{Frequency Shift in Scattering Spectrum for Different Frequency detuning}

In Fig.~\ref{fig:continuous-weak} of the main text, we have examined the steady-state superradiant Raman scattering for the systems within the crossover regime. In Fig.~\ref{fig:detunung shift}, we complement these results with the frequency shift $\Delta \omega$ (relative to the cavity mode) as function of the frequency detuning $\Delta$ of the dressed laser to the atomic transition (between the upper hyper-fine ground state and the excited state). We see that the frequency shift is negative for the negative frequency detuning ($\Delta \omega<0$ 
 for $\Delta <0$), changes the sign for the positive frequency detuning ($\Delta \omega>0$ 
 for $\Delta >0$), and tend to diverge when the frequency detuning approaches zero ($\Delta \omega\to \infty$ 
 for $\Delta \to 0$). These results seem to hint the relation $\Delta\omega \propto 1/\Delta$, and suggest that the observation might be attributed to the AC Stark shift effect.

\subsection{Continuous Superradiant Raman Scattering for Systems within Strong Coupling Regime}

\begin{figure}[!htp]
\centering
\includegraphics[scale=0.22 ]{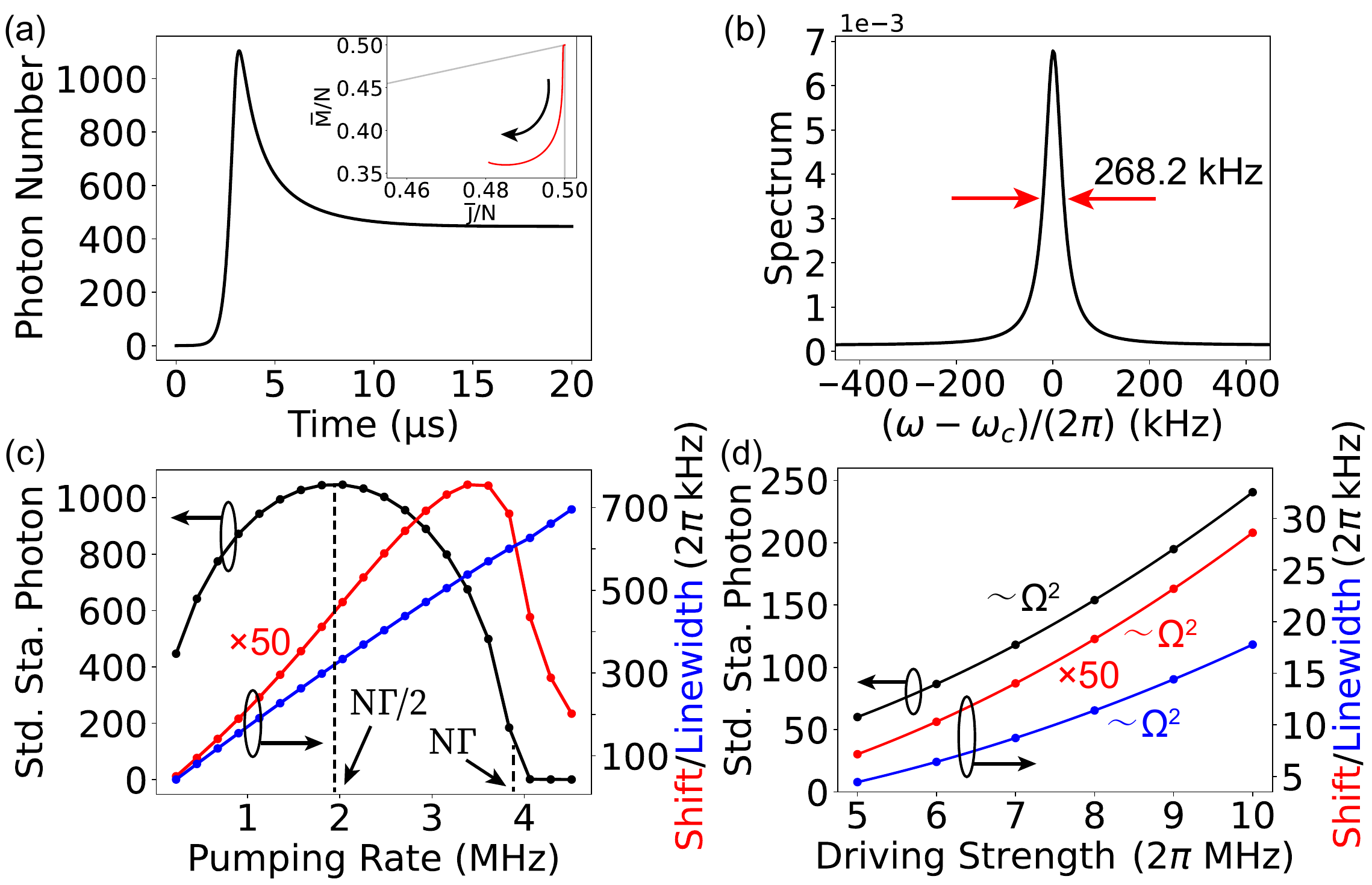}
\caption{\label{fig:continuous-strong} Continuous superradiant Raman scattering for systems with $10^6$ atoms within the strong coupling regime.  The results are similar as those in Fig.~\ref{fig:continuous-weak} for the system with $10^4$ atoms within the crossover regime, except that the steady-state photon number is much larger, the spectral shift is smaller, and the spectral linewidth is much larger. Furthermore, the intra-cavity photon number does not show the oscillations.  }
\end{figure}

Figure~\ref{fig:continuous-strong} shows the continuous superradiant Raman scattering for systems with $10^6$ atoms within the strong coupling regime and in the presence of incohernt  pumping, which are calculated with the full model. These results agree qualitatively with the those for the system within the crossover regime (Fig.~\ref{fig:continuous-weak}), except that the frequency shift is orders of magnitude less and the linewidth broadening is significantly larger. Furthermore, the intra-cavity photon number does not show any oscillations before reaching the steady-state value.

\end{document}